\documentclass[a4paper,11pt]{article}
\usepackage{pos}
\usepackage{float}
\usepackage{tabulary}
\usepackage{empheq}

\graphicspath{{../figs/}}

\def\non{\nonumber\\}

\title{Gravitational waves from supercooled phase transitions and pulsar timing array signals}

\author[a]{Jinzheng Li}
\author*[a]{Pran Nath}

\affiliation[a]{Department of Physics, Northeastern University, \\
111 Forsyth Street, Boston, MA 02115-5000, U.S.A.}

\emailAdd{li.jinzh@northeastern.edu}
\emailAdd{p.nath@northeastern.edu}

\abstract{The recent detection of a gravitational wave background in the nano-Hertz frequency range by Pulsar Timing Array (PTA) collaborations, including NANOGrav, EPTA, and PPTA, has opened a new avenue for exploring fundamental physics in the early universe. In this work, we analyze a supercooled first-order phase transition in a hidden sector with a spontaneously broken $U(1)_X$ gauge symmetry as a source for this signal.
 We demonstrate that the thermal history of the hidden and visible sectors plays a crucial role in the gravitational wave power spectrum analysis.
 Our analysis shows that supercooled phase transitions can generate gravitational waves strong enough to explain the PTA observations while satisfying cosmological constraints from Big Bang Nucleosynthesis.}

\FullConference{Proceedings of the Corfu Summer Institute 2025 "BSM 2025 Conference" (CORFU2025)\
August 25--September 3, 2025\\
Corfu, Greece\\}

\begin{document}
\maketitle

\section{Introduction} \label{sec:int}

The observation of gravitational waves from black hole mergers in 2016~\cite{Abbott:2016blz} opened a new avenue to explore fundamental physics. Among various possible sources of gravitational waves, stochastic gravitational wave backgrounds are of particular interest as they could reveal fundamental new physics including inflation, cosmic strings, and cosmic phase transitions~\cite{Caprini:2015zlo,Caprini:2019egz,Athron:2023xlk}.
Recently, evidence for a signal in the Pulsar Timing Array (PTA) in the nano-Hertz frequency range ($10^{-10}-10^{-8}$ Hz) has been found by NANOGrav~\cite{NANOGrav:2023gor}, EPTA~\cite{EPTA:2023fyk}, and PPTA~\cite{Reardon:2023gzh}. One of the most plausible explanations for this signal is a first-order phase transition (FOPT) during the early universe. The standard model of particle physics does not produce gravitational waves through a first-order phase transition, as the electroweak transition is a smooth crossover. This indicates that the signal must arise from physics beyond the standard model (BSM).

To reach the nano-Hertz PTA frequency range, a supercooled phase transition~\cite{Hawking:1981fz} is required for two reasons . First, a slow transition rate with small $\beta/H$ is needed to generate a large enough gravitational wave power spectrum. The peak amplitude scales as
\begin{align}
\Omega_{\rm GW} h^2 \sim 10^{-6} \left(\frac{H}{\beta}\right)^2 \left(\frac{\alpha}{1+\alpha}\right)^2,
\end{align}
where $\alpha$ is the transition strength parameter. Second, the phase transition should occur before Big Bang Nucleosynthesis (BBN) at $T \gtrsim 1$ MeV to satisfy cosmological constraints. The gravitational wave frequency in a FOPT is typically
 \begin{align}
  f\sim 10^{-4} \frac{\beta/H}{100} \left(\frac{T}{100 \text{ GeV}}\right) \left(\frac{g_*}{100}\right)^{1/6}
  \text{ Hz}.
  \end{align}
For non-supercooled transitions with large $\beta/H > 1000$, reaching nano-Hertz frequencies while maintaining $T \gtrsim 1$ MeV is not possible.
This note is based on ref.~\cite{Li:2025nja} which examined whether supercooled FOPTs in a hidden sector can produce gravitational waves that match the PTA observations. Several challenges arise in such analyses~\cite{Athron:2023mer}: (i) late transitions risk interfering with Big Bang Nucleosynthesis (BBN), (ii) energy release and entropy injection must not spoil light element abundances, and (iii) the transition may be too slow to complete. We demonstrate how these challenges are overcome by proper inclusion of the thermal history of the hidden and visible sectors. Similar studies on hidden sector phase transitions have been conducted in refs.~\cite{Breitbach:2018ddu,Schwaller:2015tja,Fairbairn:2019xog,Bringmann:2023opz}.

\section{A hidden sector model for gravitational wave analysis}

We consider a hidden sector model coupled to the Standard Model (SM) through a kinetic mixing portal~\cite{Holdom:1985ag}. The model consists of the SM and a hidden sector charged under a $U(1)_X$ gauge symmetry with field content: $A_\mu$ (gauge field), $q$ (Dirac fermion), $\Phi$ (complex scalar), and a kinetic energy portal. The Lagrangian is given by $\mathcal{L}= \mathcal{L}_{\text{SM}}+ \Delta\mathcal{L}$ where $\mathcal{L}_{\text{SM}}$ is the standard model Lagrangian and $\Delta\mathcal{L}$ describes the hidden sector and its coupling to the visible sector:
\begin{align}
\Delta\mathcal{L} = & -\frac{1}{4} F_{\mu\nu} F^{\mu\nu} -  \frac{\delta}{2} F_{\mu\nu} B^{\mu\nu}-|(\partial_\mu - i g_x A_\mu) \Phi|^2  \non  & + \bar{q} (i \gamma^\mu \partial_\mu - m_q) q  -  f_x \bar{q} \gamma^\mu q A_\mu - V_{0}^{\rm hid}(\Phi),
\end{align}
where $A_\mu$ is the gauge field associated with a hidden $U(1)_X$ symmetry, $\Phi$ is a complex scalar field charged under $U(1)_X$, $q$ is a dark fermion, $B_\mu$ is the SM hypercharge gauge field, and $\delta$ is the kinetic mixing parameter. The hidden sector potential is
\begin{align}
   V_{0}^{\rm hid}&= -\mu_h^2\Phi\Phi^* +\lambda_h (\Phi^*\Phi)^2,~~
\Phi= \frac{1}{\sqrt 2} (\phi_c+ \phi+ i G^0_h),
\end{align}
where $G^0_h$ is a Goldstone boson. Thermal contributions to the zero-temperature effective potential $V_{0}^{\rm hid}(\Phi)$ enable a first-order phase transition, during which the scalar field $\Phi$ acquires a vacuum expectation value (VEV). This VEV generates masses for both the hidden sector gauge boson $A_\mu$ and the scalar field $\phi$ itself. Three hidden sector fields $A_\mu, \phi, G_h^0$ contribute to the effective potential, with field-dependent masses
\begin{align}
  m_A^2(\phi_c)= g^2_x \phi_c^2,  ~~m_\phi^2(\phi_c)= -\mu_h^2+ 3 \lambda_h \phi_c^2,
  ~~m_{G^0_h}^2(\phi_c)=-\mu_h^2+  \lambda_h \phi_c^2.
\end{align}

The effective temperature-dependent hidden sector potential including loop corrections is given by~\cite{Feng:2024pab}
\begin{align}
V^{\rm h}_{{\rm eff}}(\phi_c,T_h)&=V_{0}^{\rm hid}(\phi_c)+V_{\rm 1}^{(0)}(\phi_c)
+\Delta V^{(T_h)}(\phi_c,T_h),
\label{eq:Veff}
\end{align}
where $V_{\rm 1}^{(0)}$ is the zero-temperature one-loop Coleman-Weinberg potential and $\Delta V^{(T_h)}$ is the finite thermal correction. For $V_{\rm 1}^{(0)}$ we have
\begin{align}
    V_{\rm 1}^{(0)}(\phi_c) = \sum_i \frac{g_i(-1)^{2s_i}}{64\pi^2}m_i^4(\phi_c)
  \left[\ln{\left(\frac{m_i^2(\phi_c)}{\Lambda^2}\right)}-{\cal C}_i\right],
\end{align}
where $g_i$ is the number of degrees of freedom of particle $i$, $s_i$ is its spin, and $\mathcal{C}_i = \frac{5}{6}$ ($\frac{3}{2}$) for gauge bosons (fermions). The finite thermal correction $\Delta V^{(T_h)}$ is given by
\begin{align}
    \Delta V^{(T_h)}(\phi_c,T_h) &= \frac{T_h^4}{2\pi^2}\sum_{i=\text{bosons}}g_i\int_0^\infty dq~q^2\ln\left(1-\exp{\left(-\sqrt{q^2+\mathcal{M}_i^2(\phi_c,T_h)/T_h^2}\right)}\right)\nonumber\\
    &-\frac{T_h^4}{2\pi^2}\sum_{i=\text{fermions}}g_i\int_0^\infty dq~q^2\ln\left(1+\exp{\left(-\sqrt{q^2+\mathcal{M}_i^2(\phi_c,T_h)/T_h^2}\right)}\right),
\end{align}
where $\mathcal{M}_i^2(\phi_c,T_h)$ are thermally corrected masses using Debye masses given by
\begin{align}
    \mathcal{M}_i^2(\phi_c,T_h) &= m_i^2(\phi_c) +\Pi_i(T_h), \non
    \Pi_{A}(T_h) &=  \frac{2}{3}g_x^2 T_h^2, ~~
    \Pi_{\phi}(T_h) = \frac{1}{4}g_x^2 T_h^2 + \frac{1}{3}\lambda_hT_h^2.
\end{align}

According to the model, six parameters define it: $g_x, m_q, \delta, f_x, \mu_h$ and $\lambda_h$, with an additional important parameter $\xi_0$ being the initial temperature ratio of the hidden and visible sectors, a topic discussed in further detail in Section~\ref{sec:xi}.

\begin{figure}[H]
\centering
\includegraphics[width=0.7\linewidth]{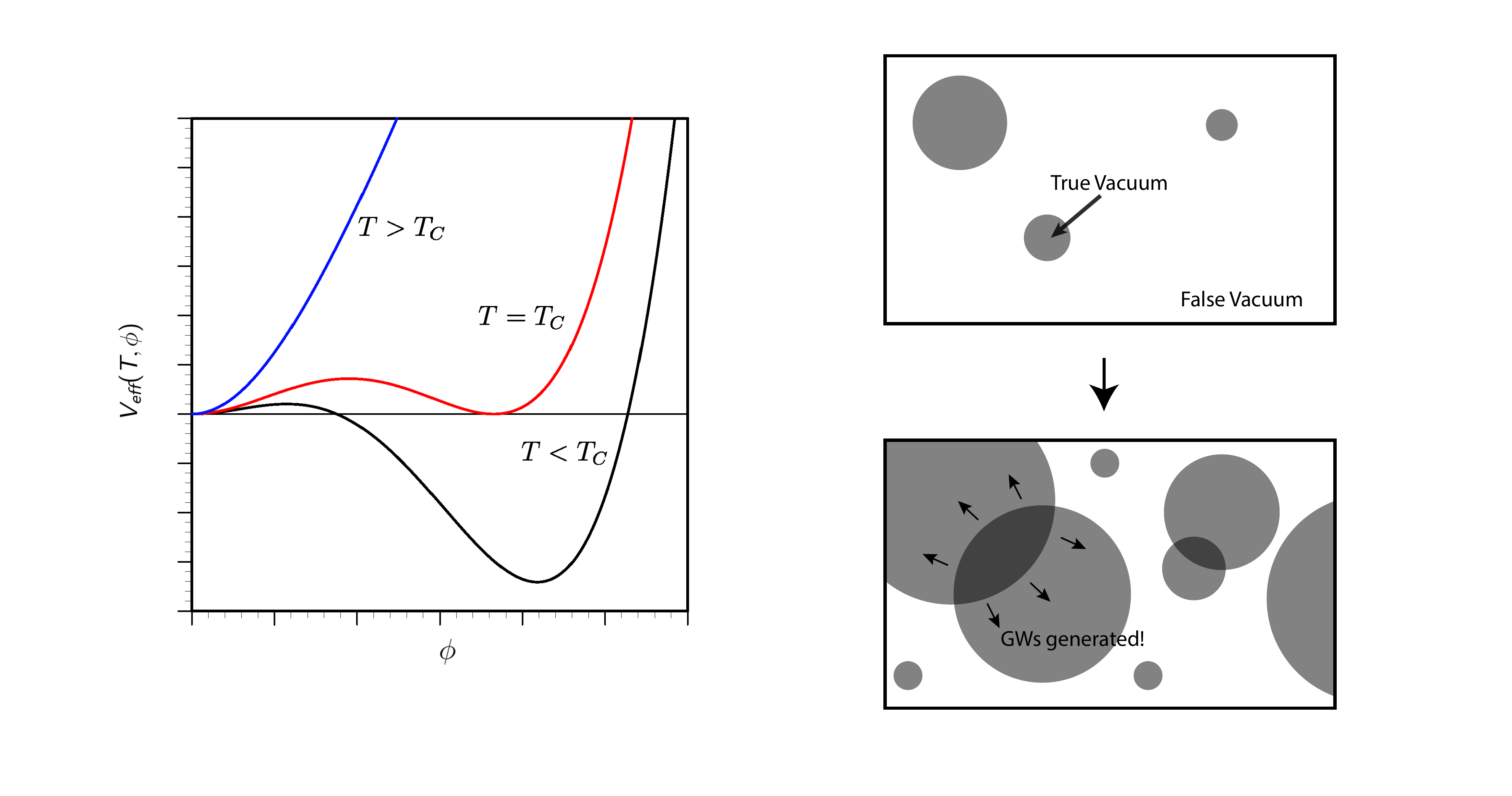}
\caption{\small Illustration of the first-order phase transition showing the temperature-dependent effective potential. At the critical temperature $T_c$, the false and true vacua are degenerate. As the temperature decreases below $T_c$, the barrier between false and true vacua enables bubble nucleation and the supercooled phase transition proceeds.}
\label{fig:FOPT}
\end{figure}

\section{Supercooled phase transition dynamics}

A supercooled phase transition occurs when the transition temperature $T_*$ is much lower than the critical temperature $T_c$, i.e., $T_*/T_c \ll 1$~\cite{Hawking:1981fz,Ellis:2018mja}. Such transitions are characterized by the transition temperature $T_*$, the transition strength $\alpha$, the inverse time scale $\beta$, the bubble wall velocity $v_w$, and the efficiency coefficient $\kappa$~\cite{Athron:2023xlk}. The parameter space can be divided into four distinct cases based on phase transition behavior~\cite{Li:2025nja}:
\begin{itemize}
\item \textbf{Case 1:} First-order phase transitions do not occur. This includes the ``ultracooled'' region where the potential barrier is too large for percolation, and the ``crossover'' region where the transition is smooth and continuous.
\item \textbf{Case 2:} Phase transitions either do not complete or the physical volume of the false vacuum does not decrease at $T_{h,p}$.
\item \textbf{Case 3:} The phase transition successfully completes with a permanent potential barrier (U-shaped action curve). This case can achieve sufficiently small transition rates to fit PTA signals.
\item \textbf{Case 4:} The phase transition completes with the metastable false vacuum disappearing at some non-zero temperature (``tangent-like'' action curve), typically yielding larger transition rates.
\end{itemize}

The percolation temperature $T_{h,p}$ should be used to describe the transition temperature $T_*$ for a supercooled phase transition. It is defined when 71\% of the universe remains in the false vacuum~\cite{Athron:2022mmm,Wang:2020jrd}:
\begin{align}
    P_f(T_{h,p}) &= 0.71,~~
         P_f(T_h)=\exp\left(-\frac{4\pi}{3}v_w^3\int_{T_h}^{T_{h,c}}dT_h'\frac{\Gamma(T_h')}{T_h'H^4(T_h')}\left(\int_{T_h}^{T_h'}dT_h''\frac{1}{H(T_h'')}\right)^3\right),
\label{eq:percolation}
\end{align}
where $\Gamma(T) \simeq T^4 \left(\frac{S(T)}{2\pi}\right)^{3/2}\exp{(-S(T))}$ is the bubble nucleation rate and $S(T) = S_3(T)/T$ with $S_3$ being the bounce action.

For supercooled phase transitions, using $\frac{\beta_*}{H_*} = T_{h} \left. \frac{dS(T_h)}{dT_h} \right|_{T_h = T_{h,p}}$ to characterize the transition rate leads to significant errors because it is just a linear coefficient in a Taylor expansion~\cite{Athron:2023xlk,Megevand:2016lpr}. Instead, the mean bubble separation $R_*$ should be used, computed directly from the bubble number density~\cite{Athron:2023mer,Enqvist:1991xw}:
\begin{align}
    R_*(T_h) = (n_b(T_h))^{-\frac{1}{3}} = \left(T_h^3\int_{T_h}^{T_{h,c}}dT_h'\frac{\Gamma(T_h')P_f(T_h')}{{T_h'}^4H(T_h')}\right)^{-\frac{1}{3}}.
\label{eq:Rstar}
\end{align}
The commonly used approximation $R_* H_*\simeq (8\pi)^{1/3} v_w/(\beta/H)$ breaks down for supercooled transitions. Using $\beta_*/H_*$ consistently yields larger values than $(8\pi)^{1/3}v_w/(R_*H_*)$, leading to underestimation of the final gravitational wave power spectrum. A large portion of parameter space that would be excluded by the condition $\beta_*/H_* > 3$~\cite{Turner:1992tz} is in fact viable when $R_*$ is used.

\section{Thermal history of hidden and visible sectors}\label{sec:xi}

\begin{figure}[H]
\centering
\includegraphics[width=0.6\linewidth]{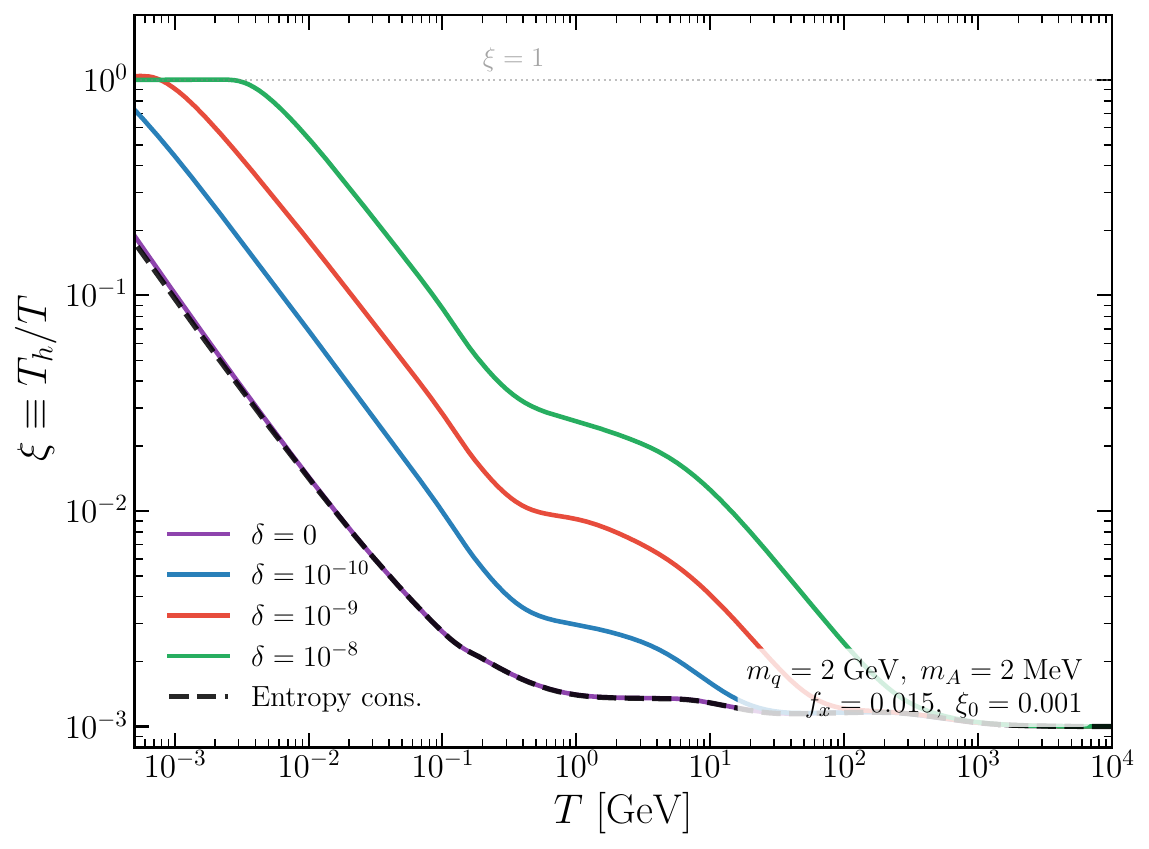}
\caption{\small Evolution of the temperature ratio $\xi \equiv T_h/T$ as a function of visible sector temperature $T$ for different values of kinetic mixing $\delta$, with $m_q = 2$~GeV, $m_A = 2$~MeV, $f_x = 0.015$, and $\xi_0 = 0.001$. The black dashed line shows the separate entropy conservation approximation. For $\delta = 0$ and $10^{-10}$, the Boltzmann solution closely follows entropy conservation. For $\delta = 10^{-9}$ and $10^{-8}$, energy transfer between sectors becomes significant, driving $\xi$ toward unity (thermalization). This demonstrates that separate entropy conservation is a good approximation only for $\delta \lesssim 10^{-10}$. Figure adapted from ref.~\cite{Li:2023nez}.}
\label{fig:xi}
\end{figure}

An important element of our analysis is the synchronous evolution of the hidden and visible sectors at different temperatures~\cite{Feng:2024pab,Aboubrahim:2022bzk}. We define the temperature ratio $\xi(T) = T_h/T$ where $T_h$ is the hidden sector temperature and $T$ is the visible sector temperature. Thermal evolution of the hidden and visible sectors are affected by couplings between the two sectors, which can exist via kinetic mixing of gauge fields~\cite{Holdom:1985ag}, Higgs portal couplings, or Stueckelberg mass mixing. Using energy conservation equations for the coupled sectors~\cite{Aboubrahim:2020lnr,Li:2023nez}:
\begin{align}
 \frac{d\rho_v}{dt} + 3 H(\rho_v+p_v)&=j_v,~~\text{(visible)},\non
 \frac{d\rho_h}{dt} + 3 H(\rho_h+p_h)&=j_h,~~\text{(hidden)},
\end{align}
where $j_v$ and $j_h$ are interaction rates between sectors, one can derive an evolution equation for $\xi(T)$. 
In the limit when the couplings between the visible and the hidden sectors vanish ($j_v = j_h = 0$),
the two sectors evolve independently and entropy is separately conserved in each sector, yielding~\cite{Li:2023nez}:
\begin{align}
    \xi(T)^3 \, \frac{h_{\rm eff}^h(\xi(T) \, T)}{h_{\rm eff}^v(T)} = \xi_0^3 \, \frac{h_{\rm eff}^h(\xi_0 \, T_0)}{h_{\rm eff}^v(T_0)} = \text{const},
\label{eq:entropy_cons}
\end{align}
where $h_{\rm eff}^{v,h}$ are the entropic effective degrees of freedom for the visible and hidden sectors, and $\xi_0$ is the initial temperature ratio at some reference temperature $T_0$. As shown in Fig.~\ref{fig:xi}, this approximation is excellent for $\delta \lesssim 10^{-10}$ but breaks down for larger kinetic mixing where energy transfer drives the sectors toward thermal equilibrium. For the benchmark points in this work with $\delta \sim 10^{-11}$, separate entropy conservation provides a reliable and computationally efficient method for tracking $\xi(T)$.

The thermal history enters the phase transition analysis in several important ways. First, the transition strength parameters depend on the temperature ratio~\cite{Bringmann:2023opz,Giese:2020znk}:
\begin{align}
    \alpha_{\rm tot} &= \frac{\Delta \Bar{\theta}(T_{h,p})}{\rho_{\rm rad}^v(T_{v,p}) + \rho_{\rm rad}^h(T_{h,p})}, ~~
    \alpha_h = \frac{\Delta \Bar{\theta}(T_{h,p})}{\rho_{\rm rad}^h(T_{h,p})},
\end{align}
where $\Delta \Bar{\theta}$ is the vacuum energy released during the transition~\cite{Giese:2020rtr}. The parameter $\alpha_{\rm tot}$ controls the gravitational wave power spectrum while $\alpha_h$ enters hydrodynamic calculations. For models with $g_{\rm eff}^v \gg g_{\rm eff}^h$, we have $\alpha_{\rm tot}\propto \xi_p^4$, which implies that 
maximizing $\alpha_{\rm tot}$ requires maximizing $\xi_p$~\cite{Breitbach:2018ddu}.
However, the effective number of additional neutrino species scales as $\Delta N_{\rm eff} \propto \xi_{\rm BBN}^4$, and BBN/CMB observations impose $\Delta N_{\rm eff} < 0.3$~\cite{Planck:2018vyg,Cyburt:2015mya}. These conditions---maximizing $\alpha_{\rm tot}$ while limiting $\Delta N_{\rm eff}$---are in conflict. This tension is resolved by considering a \textit{decaying hidden sector}~\cite{Bringmann:2023opz}: when hidden sector particles become non-relativistic and undergo ``cannibalism'' (number-changing processes like $\phi\phi\phi\to\phi\phi$)~\cite{Farina:2016llk,Ertas:2021xeh}, they consume their own particles to maintain temperature while the sector eventually decouples. This causes $g^{h,\rm rad}_{\rm eff}$ to decrease exponentially, driving $\Delta N_{\rm eff}$ to zero while allowing $\xi_p \sim \mathcal{O}(1)$ during the phase transition.

Second, the Hubble parameter $H(T_h)$ depends on the total energy density including both sectors~\cite{Li:2025nja}:
\begin{align}
    H(T_h)= \sqrt{\frac{8\pi G}{3}\left(\rho^v_{\rm rad}(T_h/\xi(T_h))+\rho^h_{\rm rad}(T_h) +\rho^h_{\rm vac}(T_h)  \right)},
\end{align}
where $\rho^v_{\rm rad}$, $\rho^h_{\rm rad}$, and $\rho^h_{\rm vac}$ are the visible radiation, hidden radiation, and hidden vacuum energy densities, respectively. Since $H(T_h)$ appears in the percolation condition Eq.~(\ref{eq:percolation}) and the mean bubble separation Eq.~(\ref{eq:Rstar}), a nontrivial evolution of $\xi(T_h)$ significantly affects the percolation temperature $T_{h,p}$ and the mean bubble separation $R_*$. The longer integration range between $T_{h,c}$ and $T_{h,p}$ in supercooled transitions amplifies this effect.

\begin{figure}[H]
\centering
\includegraphics[width=\linewidth]{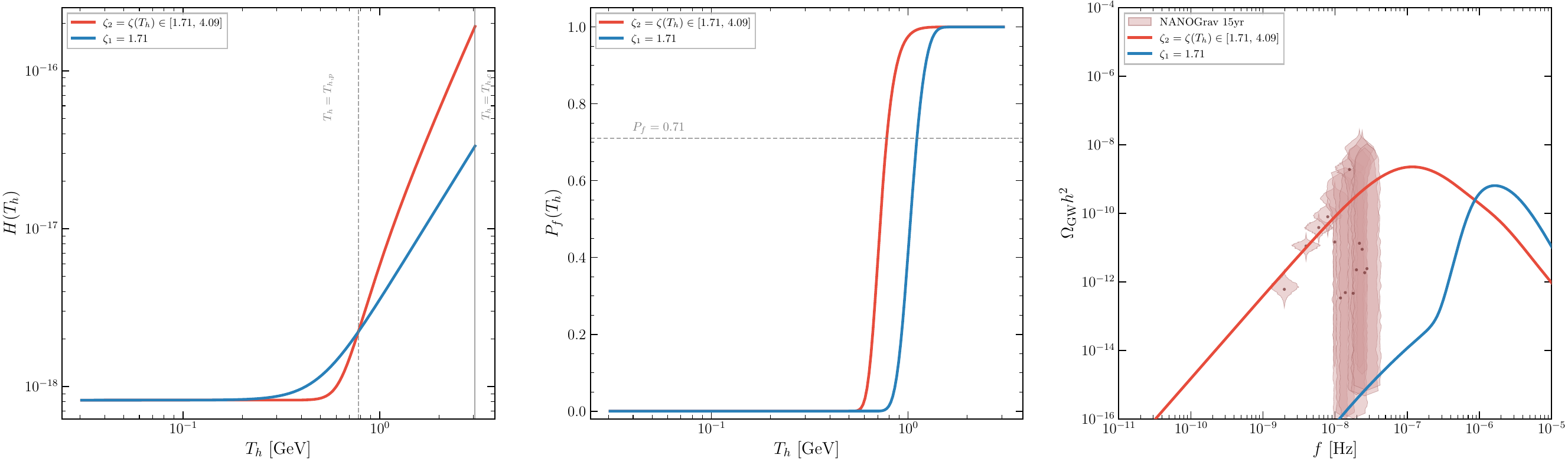}
\caption{\small Comparison of a constant vs an evolving 
temperature ratio $\zeta(T_h) = T_v/T_h$ for BP1. Left: Hubble parameter $H(T_h)$ as a function of hidden sector temperature. The dashed vertical line marks the percolation temperature $T_{h,p}$ and the solid line marks the critical temperature $T_{h,c}$. Middle: False vacuum survival probability $P_f(T_h)$; the horizontal dashed line marks $P_f = 0.71$. Right: Gravitational wave power spectrum with NANOGrav 15yr data. The red curves use the evolving $\zeta_2 = \zeta(T_h) \in [1.71,\, 4.09]$ from entropy conservation, while the blue curves use a constant $\zeta_1 = 1/\xi_p = 1.71$. Figure adapted from ref.~\cite{Li:2025nja}.}
\label{fig:zeta_comparison}
\end{figure}

Ignoring the $\xi$ evolution can lead to discrepancies of up to four orders of magnitude in the predicted gravitational wave power spectrum~\cite{Li:2025nja}. To illustrate this, we compare two choices of $\zeta(T_h) = T_v/T_h$ in Fig.~\ref{fig:zeta_comparison}: a constant value $\zeta_1 = 1/\xi_p$ (the temperature ratio at percolation) and the evolving function $\zeta_2 = \zeta(T_h)$ obtained from entropy conservation. The left panel shows that the Hubble parameter $H(T_h)$ differs by a factor of $\sim 2$--$3$ near $T_{h,c}$, since the evolving $\zeta$ accounts for the hotter visible sector at early times (when $\xi \approx \xi_0 \ll 1$). This difference in $H(T_h)$ propagates into the percolation probability $P_f(T_h)$ (middle panel), shifting the percolation temperature by $\sim 42\%$: from $T_{h,p} = 1.11$~GeV with constant $\zeta_1$ to $T_{h,p} = 0.78$~GeV with evolving $\zeta_2$. The resulting gravitational wave spectra (right panel) differ by over an order of magnitude in amplitude and exhibit a frequency shift, with only the evolving case reaching the PTA-sensitive nano-Hertz region. This demonstrates the necessity of self-consistently tracking the thermal evolution of both sectors.

\section{Gravitational wave power spectrum}

The gravitational wave power spectrum receives contributions from three sources: bubble collisions, sound waves in the plasma, and turbulence in the plasma~\cite{Caprini:2015zlo,Caprini:2019egz,Weir:2017wfa}:
\begin{align}
\Omega_{\rm GW}&\simeq \Omega_{\rm col}+\Omega_{\rm sw}+\Omega_{\rm turb}.
\end{align}
The relative importance of these contributions depends on whether the bubble expansion is in the runaway or non-runaway regime. If a true vacuum bubble generated during the phase transition continues to accelerate to a velocity close to the speed of light ($v_w \simeq 1$), this is referred to as the runaway scenario, and most of the energy is stored in the bubble walls so that gravitational waves are predominantly generated by bubble collisions. Conversely, if the bubble reaches a terminal velocity $v_w < 1$ (non-runaway scenario), most of the energy resides in the plasma and gravitational waves are primarily produced by sound waves and turbulence. The classification is made using a critical phase transition strength~\cite{Caprini:2015zlo,Bodeker:2017cim}:
\begin{align}
    \alpha_{h,\infty} =  \frac{T_{h,p}^2}{\rho^h_{\rm rad}(T_{h,p})}\left(\sum_{i=\text{bosons}}n_i\frac{\Delta m_i^2}{24} + \sum_{i=\text{fermions}}n_i\frac{\Delta m_i^2}{48}\right).
\end{align}
When $\alpha_h > \alpha_{h,\infty}$, the scenario is runaway, and the efficiency factors are given by $\kappa_\phi = 1-\alpha_{h,\infty}/\alpha_h$ and $\kappa_{\rm sw} = (\alpha_{h,\infty}/\alpha_h)\kappa(\alpha_{h,\infty},c^2_{s,f},c^2_{s,t},v_w)$, where $\kappa$ is the transition coefficient calculated using the method of refs.~\cite{Giese:2020znk,Giese:2020rtr}. The kinetic energy fraction is then
\begin{align}
    K = \frac{\alpha_{\rm tot}}{1+\alpha_{\rm tot}}\kappa(c_{s,f}^2,c_{s,t}^2,\alpha_h,v_w).
\end{align}
Our analysis shows that for supercooled phase transitions capable of producing a PTA signal, $\alpha_h$ is generally too large to yield deflagration or hybrid solutions, and all PTA events are of the detonation type with bubble wall velocity approximated by the Chapman-Jouguet velocity~\cite{Espinosa:2010hh,Ai:2023see}:
\begin{align}
v_w \approx v_J = c_{s,f}\left(\frac{1 + \sqrt{3\alpha_h(1+c_{s,f}^2(3\alpha_h-1))}}{1+3\alpha_hc_{s,f}^2}\right).
\end{align}

In the envelope approximation~\cite{Huber:2008hg,Kosowsky:1992vn}, the bubble collision contribution is
\begin{align}
h^2\Omega_{\rm col}(f) &= 1.67 \times 10^{-5}
\left(\frac{H_*}{\beta}\right)^{2}
\left(\frac{\kappa_\phi \, \alpha}{1+\alpha}\right)^{2}
\left(\frac{100}{g_*}\right)^{1/3}
\frac{0.11\, v_w^{3}}{0.42+v_w^{2}}\,
S_{\rm col}(f),
\end{align}
with spectral shape $S_{\rm col}(f) = 3.8 (f/f_{\rm col})^{2.8}/[1+2.8 (f/f_{\rm col})^{3.8}]$. We provide five benchmark points in Table~\ref{tab:benchmarks} that generate sufficiently strong gravitational waves to match the PTA signal while satisfying cosmological constraints. The resulting phase transition parameters are listed in Table~\ref{tab:results}.

\begin{table}[H]
\centering
\small
    \begin{tabular}{llllllll|ll}
  \hline
   &  $m_q$& $g_x$&$f_x$ & {$\delta$($\times10^{-12}$)}  & $\xi_0$ & $\mu_h$ & $\lambda_h$ & $m_{A}$ & $m_\phi$ \\ \hline
(BP1)&16.25&2.500&0.066&15.0&0.190&3.150&0.490&11.25&4.45\\
(BP2)&24.50&2.700&0.100&5.0&0.210&5.652&0.652&18.90&7.99\\
(BP3)&28.00&2.200&0.075&30.0&0.160&4.433&0.307&17.60&6.27\\
(BP4)&56.00&2.000&0.114&8.0&0.220&7.470&0.218&32.00&10.57\\
(BP5)&14.13&2.500&0.058&20.0&0.180&2.555&0.490&9.13&3.61\\
  \end{tabular}
        \caption{\small Benchmark points BP1-BP5 for the model parameters. All masses are in units of GeV. The scalar-gauge coupling $g_x$ determines the Coleman-Weinberg potential and the phase transition dynamics, while the fermion-gauge coupling $f_x$ controls the dark matter annihilation cross-section and relic density. The values of $f_x$ are chosen to yield $\Omega_{DM}h^2$ close to the observed value.}
   \label{tab:benchmarks}
\end{table}

\begin{table}[H]
\small
\centering
    \begin{tabular}{llllll|lllll}
  \hline
   & $T_{h,p}$ & $T_{h,reh}$ & $T_{h,c}$ & $T_{v,p}$ & $\xi_p$ & $\alpha_{tot}$& $\alpha_{h}$& $\frac{\beta_*}{H_*}$& $\frac{(8\pi)^{1/3}v_w}{H_*R_*}$ & $\Omega_{DM}h^2$   \\ \hline
(BP1)&0.778&2.021&3.053&1.330&0.585&0.284&305&-227.34&4.38&$0.114$\\
(BP2)&1.318&3.337&5.093&1.857&0.709&0.534&337&-224.38&4.20&$0.113$\\
(BP3)&0.906&3.231&4.865&1.387&0.653&1.964&2766&-316.70&3.59&$0.111$\\
(BP4)&1.959&6.009&9.008&3.023&0.648&0.852&549&-255.54&3.42&$0.110$\\
(BP5)&0.658&1.645&2.476&1.254&0.525&0.153&222&-215.79&4.14&$0.112$\\
  \end{tabular}
   \caption{\small Output parameters for benchmarks BP1-BP5. $T_{h,p}$ and $T_{v,p}$ are the percolation temperatures for the hidden and visible sectors, $T_{h,reh}$ and $T_{h,c}$ are the reheat and critical temperatures, and $\xi_p$ is the temperature ratio at percolation. All temperatures are in GeV. Note that $\beta_*/H_*$ is negative, which would exclude these points under the traditional constraint $\beta_*/H_* > 3$, but they are valid using $R_*$. All benchmarks belong to Case 3. The fermion-gauge coupling $f_x$ is tuned to yield $\Omega_{DM}h^2$ consistent with the observed value $\Omega_{DM}h^2 \simeq 0.12$~\cite{Planck:2018vyg}.}
   \label{tab:results}
\end{table}

\begin{figure}[H]
\centering
\includegraphics[width=0.49\linewidth]{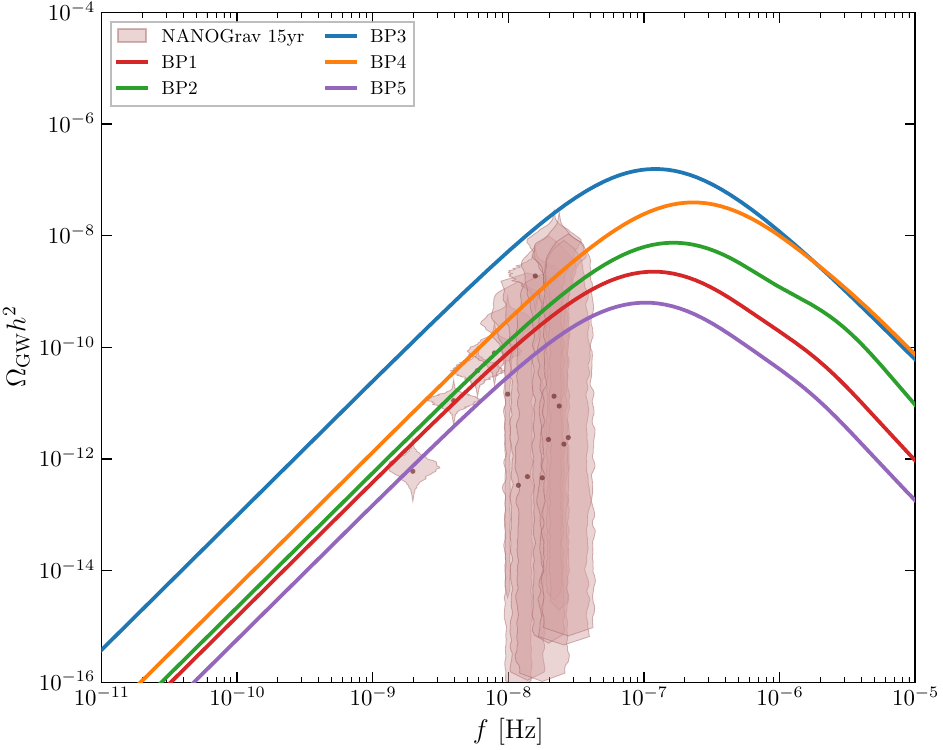}
\includegraphics[width=0.49\linewidth]{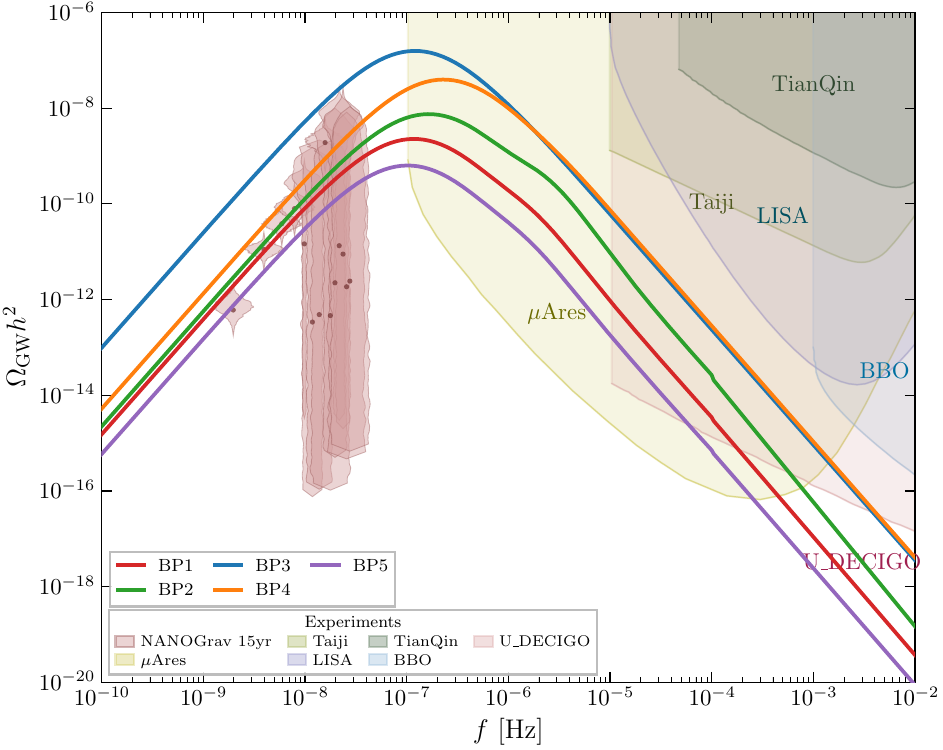}
\caption{\small Left panel: The gravitational wave power spectrum $\Omega_{\rm GW} h^2$ for supercooled phase transitions as a function of frequency for five benchmark points BP1--BP5 in the range $[10^{-11}\text{--}10^{-5}]$~Hz. The shaded violins represent the NANOGrav 15yr free-spectrum posterior~\cite{NANOGrav:2023gor}. Right panel: Projected sensitivity regions of future space-based GW detectors including $\mu$Ares~\cite{Sesana:2019vho}, Taiji~\cite{Ruan:2018tsw}, LISA~\cite{LISA:2017pwj}, TianQin~\cite{TianQin:2015yph}, BBO~\cite{Grojean:2006bp}, and U-DECIGO~\cite{Kawamura:2006up}. The benchmark spectra fall within multiple detector sensitivity regions, providing further tests of the supercooled phase transition scenario. Figure adapted from ref.~\cite{Li:2025nja}.}
\label{fig:GWcurves}
\end{figure}

\begin{figure}[H]
\centering
\includegraphics[width=0.5\linewidth]{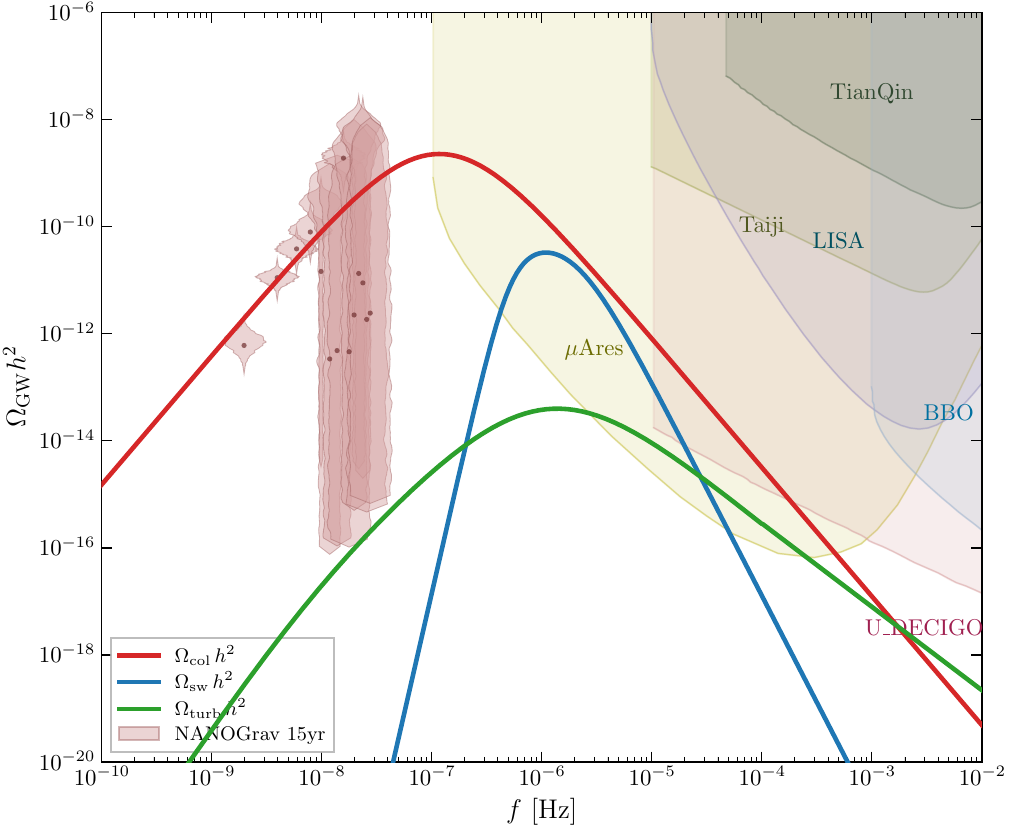}
\caption{\small The three primary contributions to the gravitational wave signal for BP1: bubble collisions (red), sound waves (blue), and turbulence (green). The bubble collision contribution dominates by several orders of magnitude in the nano-Hz PTA region, while sound waves and turbulence are negligible. The shaded regions indicate projected sensitivities of future space-based detectors. Figure adapted from ref.~\cite{Li:2025nja}.}
\label{fig:GWcontribution}
\end{figure}

In Table~\ref{tab:results}, we observe that the temperature ratio during the phase transition, $\xi_p$, ranges from $0.53$ to $0.71$, enabling strong phase transitions with $\alpha_{\rm tot}$ ranging from $0.15$ to $1.96$. For all benchmark points, the hidden sector becomes non-relativistic before BBN, satisfying $\Delta N_{\rm eff}$ constraints. Our calculations show that $\beta_*/H_*$ is negative for these benchmark points; while this would have excluded them under the constraint $\beta_*/H_* > 3$, these points are valid since we employ the mean bubble separation $R_*$ instead. For all benchmarks, we find $\alpha_{h,\infty} < \alpha_h$, confirming they are in the runaway regime.
The calculated gravitational wave power spectra for all benchmark points are displayed in Fig.~\ref{fig:GWcurves}. These spectra reach and in some cases exceed the observed PTA signal. The right panel illustrates the detection regions of future space-based gravitational wave detectors, showing that the power spectra from our benchmark points fall within these detection regions, providing further tests of the supercooled phase transition scenario.
As shown in Fig.~\ref{fig:GWcontribution}, for supercooled phase transitions generating PTA signals, bubble collisions make the dominant contribution in the nano-Hertz region, while sound waves and turbulence are highly suppressed.

\section{Conclusion}

We have demonstrated that the thermal history of the hidden sector enters crucially in the analysis of pulsar timing array signals. Taking the thermal history of the hidden versus visible sectors into account, first-order supercooled phase transitions can generate stochastic gravitational waves that explain the observations of NANOGrav, EPTA, and PPTA.
Our analysis emphasizes several key aspects: (i) the use of mean bubble separation $R_*$ rather than the inverse timescale $\beta_*$ for characterizing supercooled phase transitions---a large portion of parameter space excluded by $\beta_*/H_* > 3$ is viable when $R_*$ is used; 
(ii) the importance of synchronous evolution of hidden and visible sector temperatures;
(iii) the resolution of the tension between maximizing $\alpha_{tot}$ for PTA signals and satisfying $\Delta N_{\rm eff}$ constraints through a decaying hidden sector undergoing cannibalism.
The benchmark models presented here not only match current PTA observations but also predict signals within the detection range of future space-based gravitational wave detectors such as LISA, Taiji, TianQin, and DECIGO, providing further tests of supercooled phase transition scenarios.

\acknowledgments
The research of PN was supported in part by the NSF Grant PHY-2209903.

\end{document}